\documentclass[preprint,showpacs,preprintnumbers,nofootinbib,amsmath,amssymb]{revtex4}

\usepackage{amssymb,color}
\usepackage{graphicx}% Include figure files
\usepackage{dcolumn}% Align table columns on decimal point
\usepackage{bm}% bold math
\usepackage{subfigure}
\usepackage{hyperref}

\usepackage{color}
\usepackage[normalem]{ulem}

\def\beqn{\begin{eqnarray}} 
\def\eeqn{\end{eqnarray}} 
\def\be{\begin{equation}}
\def\ee{\end{equation}}

\begin{document}

\preprint{
\vbox{
\hbox{NPAC-12-06} 
}}

\title{Phase Transitions and Gauge Artifacts \\in an Abelian Higgs Plus Singlet Model}

\author{Carroll L. Wainwright}
\email{cwainwri@ucsc.edu} \affiliation{Department of Physics, University of California, 1156 High St., Santa Cruz, CA 95064, USA}
\author{Stefano Profumo}
%\email{profumo@scipp.ucsc.edu} 
\affiliation{Department of Physics, University of California, 1156 High St., Santa Cruz, CA 95064, USA}\affiliation{Santa Cruz Institute for Particle Physics, Santa Cruz, CA 95064, USA} 
\author{Michael J. Ramsey-Musolf}
%\email{mjrm@physics.wisc.edu} 
\affiliation{University of Wisconsin-Madison, Department of Physics
1150 University Avenue, Madison, WI 53706, USA}
\affiliation{Kellogg Radiation Laboratory, California Institute of Technology, Pasadena, CA 91125 USA}

\begin{abstract}
\noindent While the finite-temperature effective potential in a gauge theory is a gauge-dependent quantity, in several instances a  first-order phase transition can be triggered by gauge-independent terms. A particularly interesting case occurs when the potential barrier separating the broken and symmetric vacua of a spontaneously broken symmetry is produced by tree-level terms in the potential. Here, we study this scenario in a simple Abelian Higgs model, for which the gauge-invariant  potential is known, augmented with a singlet real scalar. We analyze the possible symmetry breaking patterns in the model, and illustrate in which cases gauge artifacts are expected to manifest themselves most severely. We then show that gauge artifacts can be pronounced even in the presence of a relatively large, tree-level singlet-Higgs cubic interaction. When the transition is strongly first order, these artifacts, while present, are more subtle than in the generic situation. 
\end{abstract}

%\keywords{}

%------------------------------------------------------------------------------
% User-supplied List of keywords.
\pacs{98.80.-k, 05.30.Rt, 14.80.Ec, 11.15.Ex}

% Cosmology, 98.80.-k
% quantum phase transitions, 05.30.Rt
% Neutral Higgs bosons, 14.80.Ec
%Spontaneous symmetry breaking, of gauge symmetries, 11.15.Ex

\maketitle

\section{Introduction}

Cosmological phase transitions related to the spontaneous breaking of symmetries in fundamental physics are believed to be potentially connected with the deepest questions concerning the early evolution of the universe \cite{earlypt}. 
They may pertain to such diverse topics as the origin of seed intergalactic magnetic fields (see e.g. \cite{Baym:1995fk}), the excess of baryons over anti-baryons (e.g. \cite{Trodden:1998ym}) and a possible isotropic background of gravitational waves (see e.g. \cite{Durrer:2010xc}). 
These topics are highly timely, especially in view of the recent claimed detection of small but non-vanishing intergalactic magnetic fields \cite{Neronov:1900zz, Dolag:2010ni}, progress in electric dipole moment searches \cite{Huber:2006ri, Cirigliano:2009yd}, dark matter \cite{Kozaczuk:2011vr} and direct collider searches \cite{Cohen:2012zz} for signatures of electroweak baryogenesis \cite{Cirigliano:2006dg}, and, finally, with a new generation of experiments looking for gravity waves \cite{lisapath}, that will soon boost the already significant results of current detectors \cite{ligobckg}.

The possibility of an electroweak phase transition (EWPT) associated with electroweak symmetry breaking (EWSB) is especially relevant.
In the Standard Model (SM), EWSB entails the Higgs field acquiring a non-vanishing vacuum expectation value (vev) that breaks the SU(2)$\times$U(1)$_{Y}$ gauge group down to U(1)$_{\rm e.m.}$ and generates masses of the weak gauge bosons and the SM fermions. The nature of EWSB is governed by the interplay of SM gauge interactions and the Higgs quartic self-coupling, which also determines the value of the Higgs boson mass, $m_H$.  The results of lattice simulations indicate that for $m_H \lesssim 70-80$ GeV, EWSB occurs via a first order EWPT, while for a heavier Higgs, the transition is a cross-over \cite{Kajantie:1996mn}. Given the present lower bounds on $M_H$ obtained from LEP, Tevatron, and LHC searches \cite{tevatronhiggs, lephiggs, lhcsearches}, one would conclude that an EWPT would not have occurred in a SM universe. On the other hand, extensions of the SM scalar sector can readily lead to a first order EWPT as well as associated phenomenology for collider searches.  In the context of electroweak baryogenesis the EWPT must be strongly first order in order to prevent excessive washout of the baryon asymmetry by sphaleron processes. Paradigmatic extensions to the Standard Model Higgs sector yielding a strongly first order EWPT include the minimal supersymmetric extension to the Standard Model (MSSM) with a light stop \cite{Giudice:1992hh, Carena:2008vj}, and theories (supersymmetric or not) that include one or more extra gauge-singlet fields \cite{Menon:2004wv, Huber:2006wf, Profumo:2007wc,Kang:2004pp} (other scenarios are also possible, see e.g.~\cite{Blum:2008ym}). These models typically predict distinctive collider signatures in regions of parameter space associated with a strong first order EWPT.
Consequently, rapid progress in searches for the SM Higgs at the Large Hadron Collider \cite{lhcsearches} will soon impact our understanding of a possible EWPT, elucidating whether or not the transition was strongly enough first-order for successful electroweak baryogenesis \cite{Cohen:2012zz}, whether it could have impacted thermal relic densities \cite{Wainwright:2009mq, Chung:2011hv}, and whether it could have left any detectable imprint in the diffuse background of gravitational radiation \cite{Apreda:2001us, Wainwright:2011qy}.

Once the field content of a theory is specified, the character  of resulting phase transitions relies on the computation of an effective potential, $V_\mathrm{eff}$, while its dynamics follows from the associated effective action, $S_\mathrm{eff}$. Although the most robust techniques for computing these quantities employ non-perturbative methods, such as discretizing the theory on a lattice \cite{Farakos:1994xh, Kajantie:1996mn,Aoki:1999fi}, in practice the resulting computational cost makes a perturbative calculation by far more feasible and, historically, preferentially pursued for phenomenology. However, perturbative calculations of the effective potential generically lead to gauge-dependent results, as pointed out long ago by Dolan and Jackiw \cite{jackdolan}
(see also Ref.~\cite{Weinberg:1974hy} for early work on gauge-dependence and symmetries at high temperature).
Although it is straightforward to maintain gauge-invariance when carrying out calculations in the symmetric phase (where the vev is gauge-independent), such as computing the temperature of a second-order phase transition \cite{Ueda:1980rx, Arnold:1991uh, Kelly:1994ew}, it is doing so for the broken phase requires additional care. 
The generic dependence of the effective action on the gauge choice is described by  the so-called Nielsen identities \cite{nielsen} and their generalizations \cite{followupnielsen}. In practice, the gauge invariance of the effective action is guaranteed when the background field $\varphi(x)$ is an extremal configuration, i.e. one that satisfies the equations of motion (in the case of the effective potential, $\varphi(x)=\varphi_{\rm min}$ is the value of the field corresponding to a minimum of the potential)\footnote{More generally, all background fields must be in extremal configurations. For example, the electroweak sphaleron involves non-vanishing scalar and gauge fields at the saddle point of the effective action.}. Typically, gauge dependence stems from an inconsistent truncation of the perturbative expansion \cite{mjrmpatel}. This leads, in turn, to effects in the critical temperature \cite{jackdolan2},  in the bubble nucleation rate \cite{bubbles}, and in the sphaleron transition rate \cite{mjrmpatel} for a first order phase transition, ultimately resulting in unphysical gauge dependence in observable quantities such as the spectrum of gravity waves produced by bubble collision or turbulence \cite{Wainwright:2011qy}. 
 
The problem of gauge dependence as it relates to the description of cosmological phase transitions has recently attracted renewed attention. Following earlier studies \cite{Laine:1994zq,Baacke:1993aj,Baacke:1994ix}, Ref.~\cite{mjrmpatel} addressed the possibility of producing a consistent, order-by-order gauge-invariant result in the perturbative expansion of the $V_\mathrm{eff}$ and $S_\mathrm{eff}$. In the SM, this approach yields a gauge-invariant critical temperature $T_C$ at one-loop order in a straightforward manner, and it can be refined to reproduce leading terms in the \lq\lq daisy resummation" in a gauge invariant manner. Doing so reproduces trends with model parameters that are observed in lattice studies, including the dependence of $T_C$ on the top squark soft mass parameters in the MSSM. Application to computation of a gauge invariant sphaleron rate is also feasible. On the other hand, a gauge-invariant computation of the bubble nucleation rate in the SM requires going beyond one-loop order. 

As an alternative to the SM, the Abelian Higgs model provides a theoretically attractive \lq\lq laboratory" in which to assess various approaches to obtaining gauge-invariant quantities associated with symmetry-breaking. Apart from calculational ease, this model allows for the computation of a gauge-invariant effective potential using a Hamiltonian approach~\cite{boyan}, making a comparison with results obtained with other methods possible. In Ref.~\cite{Wainwright:2011qy}, we explicitly addressed the case of an Abelian Higgs model, and calculated for a full set of $R_\xi$ gauge choices the impact of gauge-dependence on physical observables. Doing so allowed us to directly compare the results of the computation for a generic $R_\xi$ gauge choice with the gauge-independent calculation \cite{fischler, boyan}.

While computationally tractable, the Abelian Higgs model arguably carries limited phenomenological interest. Among the more relevant SM extensions   mentioned above -- such as MSSM with a light stop or a gauge-singlet extension to the Higgs sector -- it is possible to make the electroweak phase transition strongly first order via interactions that are gauge independent (though the full $V_\mathrm{eff}$ remains gauge-dependent). The simplest cases involve introduction of  tree-levels term in the potential of the type $SH^\dagger H$ or $S^2 H^\dagger H$ in the case of the real singlet extension. The tree-level cubic operator can produce a large potential barrier between the broken and unbroken phase at the electroweak phase transition, while the quartic interaction may allow for a lowering of $T_C$ in a manner compatible with  collider constraints on $m_H$. As these operators are manifestly gauge-invariant, one may inquire as to whether perturbative computations of the EWPT properties in the associated models are quantitatively less susceptible to gauge-dependent artifacts than in either the SM or in the Abelian Higgs model. Indeed, Refs. \cite{Espinosa:2011ax, Cohen:2012zz} have recently suggested that such a situation may occur. 

In what follows, we study the  issues described above in some detail. For the sake of comparing with a known, simple gauge-independent result, we shall again use the Abelian Higgs model, supplemented with a gauge-singlet real scalar field (which does not impact the gauge-dependence structure of the theory) and retaining only the tree-level cubic operator,  $SH^\dagger H$. Arguably, this model is the simplest prototypical electroweak-like theory that can exhibit a strongly first order phase transition driven by tree-level cubic terms. In view of the recent LHC results pointing to a relatively heavy Higgs mass, singlet extensions to the electroweak scalar sector have additionally become phenomenologically more appealing, making an assessment of the gauge artifacts in the effective potential even more timely. We show that such gauge artifacts may arise even in the presence of a large tree-level singlet-Higgs cubic coupling. However, we also find that the gauge-dependence is less pronounced when the tree-level and loop-induced cubic interactions conspire to generate a sizeable barrier between the broken and unbroken phases at low temperatures. 

The remainder of this paper is organized as follows: the next section describes in detail the theory we study, including explicit calculations of the gauge dependent terms in the effective potential. The following Section \ref{sec:patterns} gives an outline of the possible patterns of spontaneous symmetry breaking, and describes the effects of gauge choices on various quantities of interest (including the critical temperature, the latent heat, and a measure of the strength of the phase transition). Finally, Section \ref{sec:concl} summarizes and concludes.

\section{The Abelian Higgs Model plus a Singlet Scalar}\label{sec:model}

We  examine the gauge dependence of a simple Abelian Higgs model containing a single complex scalar $\Phi$ charged under a local U(1) gauge group, and a real scalar singlet field $s$. For our purposes here, we consider only a cubic coupling between the two fields, as the latter can generate a tree-level barrier between the broken and unbroken phases and, thus, 
 can increase the strength of the phase transition. The effect of the quartic operator discussed above is more subtle, so for simplicity we focus on the cubic interaction.  At tree-level, the potential is
\begin{equation}
V_0(\Phi, s) = \tfrac{1}{4}\lambda_1(\Phi^\dagger \Phi)^2 + \tfrac{1}{2}\mu_1^2\Phi^\dagger\Phi + \tfrac{1}{4}\lambda_2 s^4 + \tfrac{1}{2}\mu_2^2 s^2 + \tfrac{1}{2}Es\Phi^\dagger\Phi.
\end{equation}
It is useful to separate $\Phi$ into real and imaginary parts ($\Phi = h+ih'$) and then rotate into a basis such that only the real part gets a vev. 

Rather than specifying the 5 coefficients explicitly, we find it more convenient to specify the vevs, the tree-level mass eigenstates, and the mixing of the mass eigenstates, and use this to set the tree-level potential. The tree-level mass-squared matrix is
\begin{align}
\label{eq:treemass}
M^2_{ij} &= \begin{pmatrix}
3\lambda_1 h^2 + \mu_1^2 + Es   &   Eh \\
Eh   &   3\lambda_2 s^2 + \mu_2^2
\end{pmatrix} \\
&= \begin{pmatrix}
\cos \theta & -\sin\theta \\ \sin\theta & \cos\theta
\end{pmatrix}
\begin{pmatrix}
m_1^2 &0 \\ 0& m_2^2
\end{pmatrix}
\begin{pmatrix}
\cos \theta & \sin\theta \\ -\sin\theta & \cos\theta
\end{pmatrix}\ \ \ ,
\end{align}
where the rotation angle $\theta$ gives the mass eigenstate mixing:
\be
\tan (2\theta) = \frac{2 M_{12}^2}{M_{11}^2-M_{22}^2}.
\ee
The quantities $m_1$, $m_2$, and $\theta$ are most relevant to collider phenomenology, as they determine production cross sections and decay branching ratios.  In particular, a non-zero mixing angle $\theta$ can weaken the collider constraints on the lightest mass eigenstate since its effective coupling to gauge bosons is reduced by $\cos\theta$. This effect opens up the possibility of a (stronger) first order phase transition by allowing for a smaller Higgs quartic self-coupling,  although we do not study this effect in detail here.
The corresponding minimization conditions are given by
\begin{align}
\label{eq:dVdh}
\frac{\partial V_0}{\partial h} &= (\lambda_1 h^2 + \mu_1^2 + Es) h = 0, \\
\label{eq:dVds}
\frac{\partial V_0}{\partial s} &= \lambda_2 s^3 + \mu_2^2 s + \tfrac{1}{2}Eh^2 = 0.
\end{align}
From these we solve for  $\lambda_1, \lambda_2, \mu_1^2, \mu_2^2$ and $E$ in terms of vacuum expectation values of $h$ and $s$, masses  $m_1$ and $m_2$, and the angle $\theta$, assuming that both $h$ and $s$ are non-zero. Note that for $E<0$ (so that the $\langle s\rangle > 0$) and $m_1 < m_2$, we require $0 \leq \theta < 90^\circ$. Also, $\lambda_2$ is negative for sufficiently large $E$, so not all values of $\theta$ lead to stable potentials.

\begin{figure}[t] %  figure placement: here, top, bottom, or page
   \centering
   \includegraphics[scale=1]{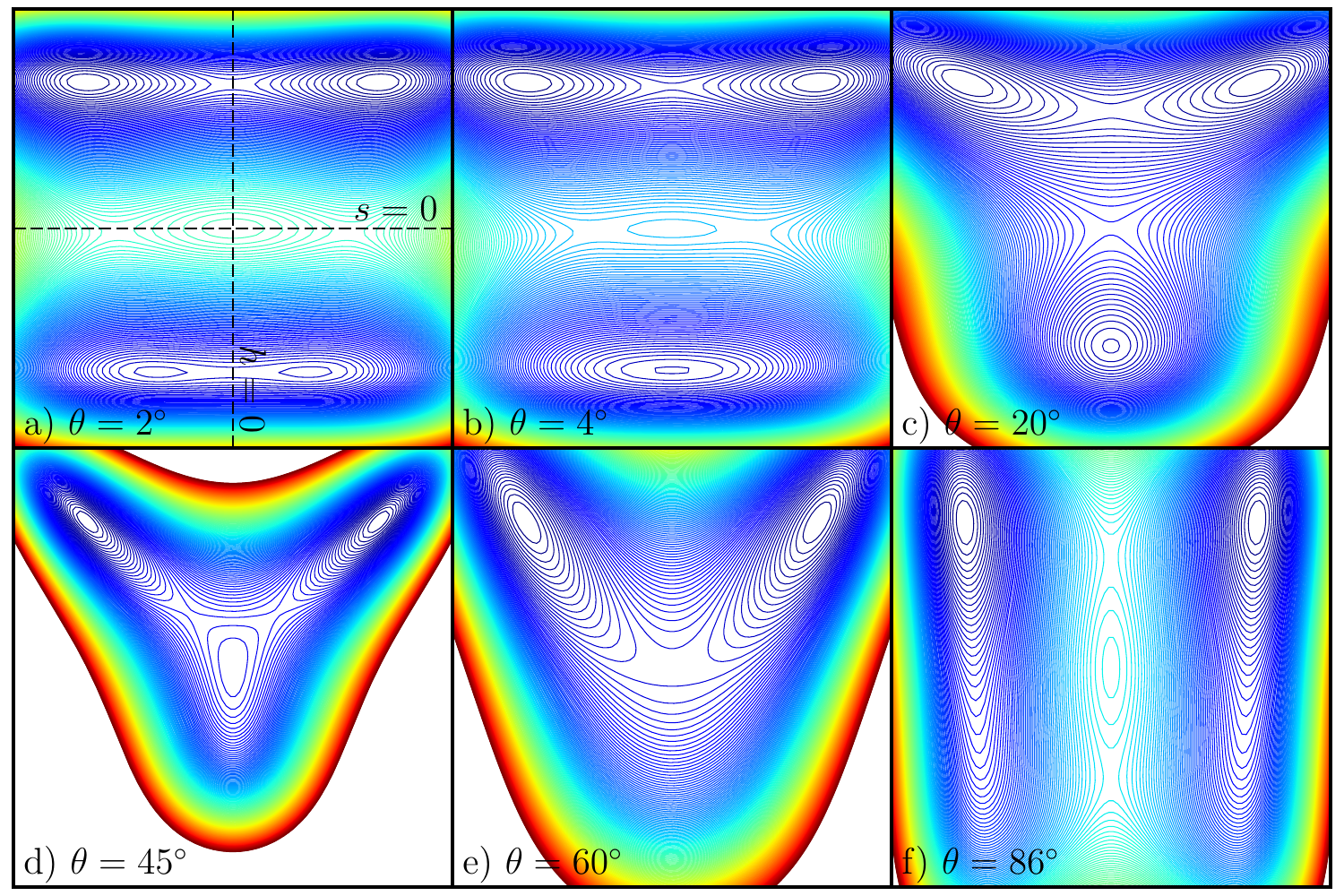} 
   \caption{Contours of the tree-level potential for $m_1/m_2 = 0.4$ and six different values of $\theta$. Red (blue) contour lines denote higher (lower) values of the potential. The Higgs and singlet fields vary along the horizontal and vertical axes, respectively. The origin, which is in the center of each plot, is a maximum in (a), (b), and (f), a saddle point in (c) and (e), and a minimum in (d).}
   \label{fig:treepot}
\end{figure}

For $h=0$, extrema occur at $s=0$ and --- for $\mu_2^2<0$ --- at $s=\pm |\mu_2| /\sqrt{\lambda_2}$. An additional three extrema can occur for $h>0$, whose locations are trivially determined by solving the cubic equation in $s$ obtained from  from combining Eqs.~(\ref{eq:dVdh}) and (\ref{eq:dVds}):
 \begin{equation}
 \label{eq:scubed}
 \lambda_2 s^3 + (\mu_2^2 - \frac{E^2}{2\lambda_1})s - \frac{E}{2\lambda_1}=0.
 \end{equation}
Since Eq.~(\ref{eq:scubed}) has no term proportional to $s^2$, at most two of the extrema can be in any one quadrant. Also, since there are no linear or cubic terms in $h$, all maxima must lie along the $s$-axis. Using this knowledge, one can enumerate all of the different combinations of minima, maxima, and saddle points to obtain all of the different possible potential types.

From Eqs.~(\ref{eq:dVdh},\ref{eq:dVds}) we observe that the location and character of the extrema depend on four independent parameters. For example, by scaling out a factor of $\lambda_1$ we may take these parameters to be $\mu_1^2/\lambda_1$,  $\mu_2^2/\lambda_1$, $\lambda_2/\lambda_1$, and $E/\lambda_1$. We may trade two of these parameters for one each of the non-zero vacuum expectation values of $s$ and $h$, respectively. The remaining two parameters then determine $\theta$ and the ratio of masses $m_1/m_2$. The depth of the potential at one of the minima is then fixed by the fifth remaining parameter in the potential, which we can trade off for one of the masses. Therefore, we can keep two vevs and one of the masses fixed  and just vary $\theta$ and the ratio $m_1/m_2$ to explore all potentials that are not related by an overall rescaling. Fig.~\ref{fig:treepot} shows six different representative potentials with constant $m_1/m_2$.

Panel (\emph{a}) shows the potential with a small positive value of $\theta$, corresponding to a small negative $E$. There is a minimum in each quadrant of the $h$-$s$ plane separated by four saddle points with a maximum at the origin. Increasing the angle $\theta$ (panels (\emph{b}) and (\emph{c})) merges some of these features onto the $s$-axis. First, the two minima at $s < 0$ merge, then the
two saddle points near $s=0$ merge onto the origin such that the origin is no longer a maximum. Increasing $\theta$ further (panel (\emph{d})), pushes the minimum along the $s$-axis up to the origin. At this point there is a tree-level barrier between the broken and symmetric phases. In panel (\emph{e}), this barrier disappears and there are only the two electroweak minima and a saddle point at the origin. Finally, panel (\emph{f}) has a small value of $|E|$, but large enough to destroy the two metastable minima in panel (\emph{a}). A further rotation to $\theta=90^\circ$ would reproduce a potential of a type similar to what shown in panel (\emph{a}). This basically exhausts all of the possibilities with symmetry-breaking minima: the only other potential type occurs at larger ratios of $m_2/m_1$. It is similar to type (\emph{d}) except that the $h=0$ minimum splits into two minima and a saddle point along the $s$-axis.

\subsection{Quantum corrections}

The one-loop corrections to the effective potential are given by
\begin{equation}
\label{eq:V1}
V_1 = \sum_i \frac{n_i}{64\pi^2} m_i^4 \left(\log \frac{m_i^2}{q^2} - c\right),
\end{equation}
where $n_i$ is the d.o.f. for each particle, $m_i$ is the particle mass, $q$ is the renormalization scale (which we set to 1 TeV), and $c=1/2$ for transverse gauge boson polarizations and $3/2$ for all other particles. 
The Higgs and scalar masses are given by the eigenvalues of the tree-level mass matrix (Eq.~\ref{eq:treemass}). The three gauge boson polarizations each contribute a mass $m_{gauge}^2 = g^2h^2$. We focus on $R_\xi$ gauge\footnote{
In a previous study~\cite{Wainwright:2011qy}, we compared $R_\xi$ gauge to the gauge-independent potential in Ref.~\cite{boyan} which uses a Hamiltonian formalism. The latter is more computationally intensive, and it tends to closely resemble Landau gauge ($\xi=0$) in the Abelian Higgs model.
}, in which there are gauge-dependent masses for the Goldstone boson and the ghost: $m_{gold}^2 = \lambda_1 h^2 + \mu_1^2 + Es + \xi g^2h^2$ and $m_{ghost}^2 = \xi g^2h^2$, with $n_{ghost}=-2$.
There is an additional degree of freedom from the gauge boson's unphysical time-like polarization which exactly cancels one ghost degree of freedom.
In all of our models the first derivatives (${\partial V_1}/{\partial h}$ and ${\partial V_1}/{\partial s}$) are negative at the tree-level minimum. This pushes the vevs further away from the origin than they were at tree-level.

At finite temperature, the one-loop corrections are
\begin{equation}
V_{1,T\neq0} = \frac{T^4}{2\pi^2} \sum_i n_i J\left[\frac{m_i^2(\phi)}{T^2}\right],
\end{equation}
where
\begin{equation}
J(x^2) \equiv \int_0^\infty dy\; y^2 \log \left(1- e^{-\sqrt{y^2+x^2}}\right).
\end{equation}
In the high-temperature (low-$x$) limit,
\begin{equation}
\label{eq:high}
J(x^2) \approx -\frac{\pi^4}{45} + \frac{\pi^2}{12}x^2 - \frac{\pi}{6}x^3 - \frac{x^4}{32}\log\frac{x^2}{a_b} - \mathcal{O}(x^6)
\end{equation}
where $\log a_b = \tfrac{3}{2} - 2\gamma_E + 2\log(4\pi)$ and $\gamma_E$ is the Euler constant \cite{Anderson:1992}. $J(x^2)$ is not analytic at $x^2=0$, and it is complex for $x^2<0$. In our calculations, we approximate $J(x^2)$ with a cubic spline, taking only the real component for $x^2<0$.

At high temperature, the validity of the perturbative expansion of the effective potential breaks down. Quadratically divergent contributions from non-zero Matsubara modes must be re-summed through inclusion of thermal masses in the one-loop propagators \cite{gross1981,parwani1992}: $m^2(\phi) \rightarrow m^2_{eff}(\phi) = m^2(\phi)+m^2_{therm}(T)$.
This amounts to adding thermal masses to the scalars and gauge boson longitudinal polarizations:
\begin{align}
M^2_{ij} & \rightarrow M^2_{ij} + T^2\begin{pmatrix}\lambda_1/3+g^2/4 &0 \\0 & \lambda_2/4\end{pmatrix} \\
m^2_{gold} &\rightarrow m^2_{gold} + T^2(\lambda_1/3+g^2/4) \\
m^2_{long-gauge} &\rightarrow m^2_{gauge} + T^2g^2/3.
\end{align}
Note that the coefficients of the $T^2$ terms are $\xi$-independent. 

When the phase transition is second-order or very weakly first-order, or when the temperature is very high, even the re-summed potential may not be reliable. Loops that are either infrared divergent or dominated by the infrared regime contribute linearly in temperature and can ruin the perturbative expansion (see, e.g., Ref~\cite{Arnold:1992rz}). Each additional such loop contributes roughly $\tilde{\lambda} T/M$, where $\tilde{\lambda}$ is the relevant coupling and $M$ is the relevant mass scale. Substituting the gauge boson mass for $M$ and $g^2$ for $\tilde{\lambda}$, we see that the perturbative expansion should hold as long as $h/T \gtrsim g$. We warn the Reader that for certain parameters in what follows this criterion breaks down. As a result, the one-loop expansion might have limited validity in those cases. This is particularly true for cases 1 and 2 with $g=0.5$ and case 3 with $\theta \gtrsim 70^\circ$. However, it is important to note that one could lower the phase transition temperature by e.g. extending the gauge group and adding extra gauge bosons. Extra degrees of freedom enhance the finite-temperature contributions relative to the tree-level potential, so symmetry breaking happens at lower temperatures. This in turn would make the one-loop perturbative expansion more reliable, without qualitatively changing the nature of the explicit gauge dependence. Since the appearance of the one-loop gauge dependence is not tied directly to the perturbative validity, and since our primary interest is in providing proof of existence for the gauge-dependence issues, we leave the perturbative breakdown problem to future studies.

%This is true for most of the cases that we study below ({\bf this is no longer really true}), including cases that show strong patterns of gauge-dependence. These trends continue towards more weakly first-order transitions where the gauge-dependence tends to be more severe and the expansion less reliable.

\section{Patterns of spontaneous symmetry breaking}\label{sec:patterns}

We now study the gauge dependence in a few representative models with different qualitative features. In particular, we examine the gauge dependence of the nucleation temperature $T_*$  and various measures of the phase transition strength. Nucleation occurs when the three-dimensional action $S_3$ of a nucleated bubble satisfies $S_3/T_*\approx 140$ (see Refs.~\cite{coleman, linde} for original work on cosmological phase transitions), and we use this criterion to define $T_*$. The phase transition strength has often been characterized by $\Delta\phi$, the jump between the vevs of the two phases. However, as noted above, this quantity is $\xi$-dependent. Alternate, physically meaningful measures include (a) $\alpha$, the difference in energy densities at the two vevs , and (b)  a measure of the phase transition duration $\beta^{-1}$, defined as $\beta/H_* = T_* (d/dT)(S_3/T)$ where $H_*$ is the Hubble parameter at the time of the transition. When the phases are degenerate and there is no supercooling, $\alpha$ is equivalent to the transition's latent heat. The quantity $\beta^{-1}$ effectively measures the strength of the transition, with $\beta = \infty$ for a second-order transition. 

We use the {\rm CosmoTransitions} package \cite{Wainwright:2011b} to determine the phase structures and calculate the nucleation rates.

\begin{table}[h]
\caption{Model parameters for three illustrative cases.}
\begin{center}
\begin{tabular}{r|cccccc}
&$\langle h_{tree}\rangle$ & $\langle s_{tree}\rangle$ & $m_1$ & $m_2$ & $\theta$ & $E$ \\
\hline
Case 1 & 240 GeV & 48 GeV & 120 GeV & 35 GeV & $-1^\circ$ & $-0.96$ GeV \\
Case 2 & 240 GeV & 48 GeV & 120 GeV & 180 GeV & $45^\circ$ & $-37.5$ GeV \\
\hline
Case 3 & 174 GeV & 174 GeV & 30 GeV & 75 GeV & $60^\circ$ & $-11.8$ GeV \\
\end{tabular}
\end{center}
\label{tab:params}
\end{table}%

\subsection{Gauge fields driving transitions}
\begin{figure}[t]
\centering
\subfigure{
\includegraphics[scale=1]{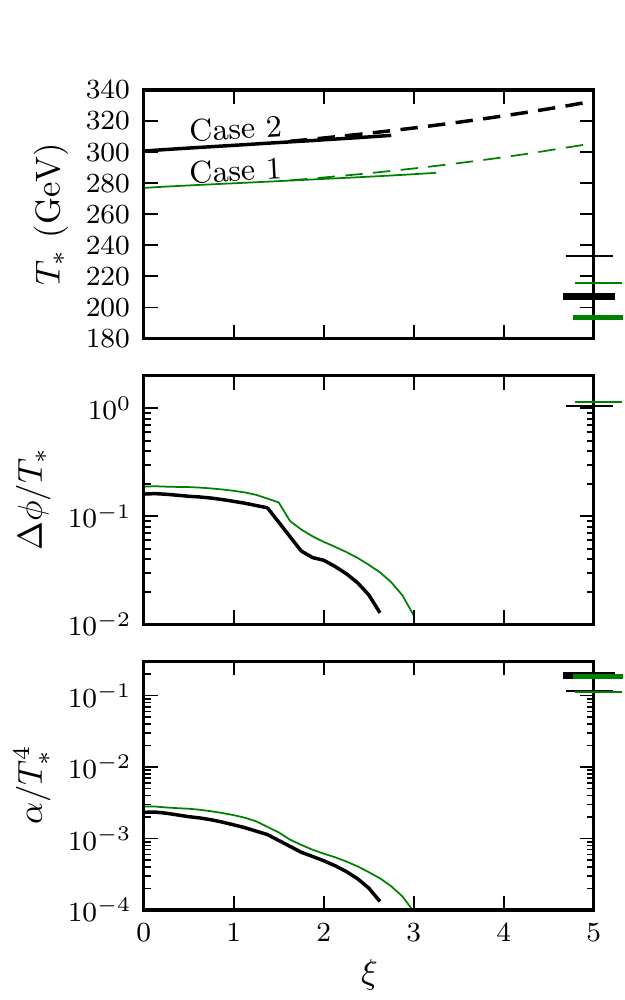}
}
\subfigure{
\includegraphics[scale=1]{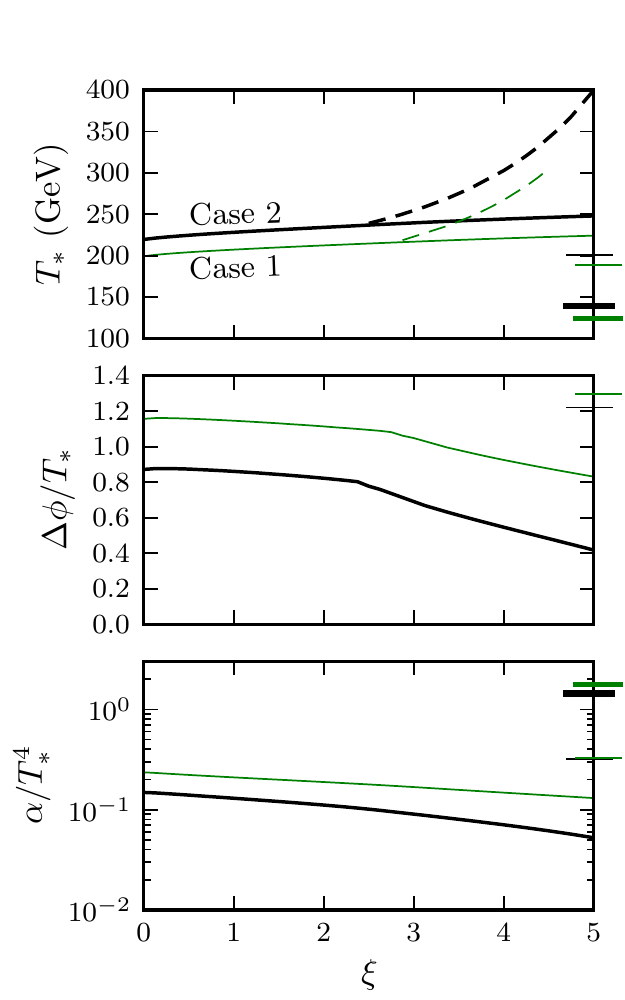}
}
\caption{\label{fig:varxi}
Gauge dependence in cases 1 (thin green lines) and 2 (black lines) 
as a function of the gauge parameter $\xi$.
The left panels have a small gauge coupling $g=0.5$, while the right have $g=1.0$. Dashed lines represent second-order symmetry breaking transitions, which may be followed by a first-order transition at lower temperature. The marks along the right side of each panel show the corresponding quantity calculated using the gauge-invariant method of Ref.~\cite{mjrmpatel}. The thicker marks include a gauge-invariant treatment of the thermal masses; the thin marks ignore them. }
%Lines with different dashes correspond to separate phase transitions within the same models. The thick marks along the right side of each panel show the corresponding quantity calculated using the gauge-invariant method of Ref.~\cite{mjrmpatel}. }
\end{figure}

First, we examine two cases in which the transition would be second-order without the inclusion of massive gauge fields. Table~\ref{tab:params} contains the corresponding model parameters. The two cases are quite similar: they have the same vevs, and they both have a saddle point at $h=s=0$ with no other extrema along the $s$-axis (see Fig.~\ref{fig:treepot}(\emph{e})). However, case 1 has a very small cubic term while in the second case $E$ is of the same order as the mass scale of the theory. The gauge dependence exhibited in case 1 is therefore unsurprising: since it has relatively weak coupling between the Higgs and the scalar singlet, we expect it to show the same sort of gauge dependence as an uncoupled Abelian Higgs model \cite{Wainwright:2011qy}. This is indeed the case, as seen in Fig.~\ref{fig:varxi} (green lines).

For low values of the gauge coupling $g$, the gauge dependence is quite pronounced. The initial symmetry breaking (solid lines for the two cases) is weakly first-order ({\em e.g.}, $\alpha/T_*^4 \ll 1$) from $\xi=0$ to $\xi \approx 1.5$. At higher $\xi$, a second-order transition initially breaks the symmetry (dashed lines in Fig.~\ref{fig:varxi}). The first-order transition then proceeds from a high-temperature broken phase to the low-temperature broken phase at larger values of $h$ and $s$. Above $\xi \approx 3$, the barrier between the two broken phases disappears and there is no first-order phase transition at all.

%For low values of the gauge coupling $g$, the gauge dependence is quite pronounced. The initial symmetry breaking (solid lines for the two cases) is weakly first-order at $\xi=0$ ({\em e.g.}, $\alpha/T_*^4 \ll 1$), and it becomes second order at $\xi \approx 1$ (where the latent heat effectively vanishes, corresponding to the endpoint of the solid lines). The subsequent phase transitions (dashed lines) start out weakly first-order at low $\xi$ but become strongly first-order as $\xi$ increases. They merge into a single first-order phase transition at $\xi\approx 0.75$. These lower-temperature, low-$\xi$ transitions appear to be due to negative mass-squared values and the non-analyticity of $J(x)$ at $x=0$, leading to kinks (discontinuous first derivatives) and extra minima in the potential. 

For higher values of $g$, the gauge dependence is not as pronounced. The phase transition is much more strongly first-order, with $\alpha/T_*^4$ roughly a factor of 10 higher than it is for low $g$. The initial symmetry-breaking still turns second-order at high $\xi$, but the subsequent first-order transition persists up to $\xi=5$ with about a factor of 2 drop in $\alpha$.

%For higher values of $g$, the gauge dependence is not as pronounced. The phase transition is strongly first-order, with $\alpha/T_*^4$ of order unity. 

In all cases, the appearance of a strong first-order phase transition is associated with a large ratio of field-dependent heavy degrees of freedom (in this case, gauge bosons) to the Higgs mass. A small Higgs mass decreases the depth of potential (that is, $V_0(0) - V_0(v)$ decreases), while heavy additional field-dependent masses yield larger contributions to the thermal effective potential (increasing $V_1(v,T)-V_1(0,T)$). Both effects decrease the critical temperature and increase $\Delta\phi$. 
The presence of the additional field-dependent masses decreases the critical temperature because, for a given value of $T$, $V_1(v,T)-V_1(0,T)$ is larger for larger gauge couplings and the two minima are degenerate at lower temperatures.  It increases the value of $\Delta\phi$ because ${dJ}/{dx} \rightarrow 0^+$ as $x \rightarrow \infty$, so when $x = m/T = gh/T$ is large, $\partial V_1/\partial h$ is small and the vev does not decrease much from its tree-level value. An increase in $\Delta\phi$ tends to increase both $\alpha$ (a larger separation between phases implies a larger difference in mass spectrums, entropy, and therefore latent heat) and $\beta^{-1}$ (since $S_3$ scales as $(\Delta\phi)^3$).
One can achieve a strongly first-order phase transition even for a heavy Higgs, as long as it is somewhat light compared to the other field-dependent masses.

Interestingly, case 2 (thick black lines) shows almost exactly the same gauge dependence as case 1, even though it has a non-trivial cubic term. The important point is that, although large, the cubic term is not large enough to cause a first-order phase transition without additional bosons that have large couplings to the Higgs field.

We compare the gauge-dependent calculations to explicitly gauge-independent calculations (denoted by marks along the right side of each panel in Fig.~\ref{fig:varxi}). At one-loop order and without the added thermal masses, the gauge-independent calculation is simply the value of the potential evaluated at the tree-level minimum $\Phi_0$, where ghost and goldstone degrees of freedom exactly cancel. Thermal masses spoil the cancellation, but one can still obtain a gauge-invariant result by evaluating the cubic terms in the ring-improved effective potential at the tree-level high-temperature minimum, where the tree-level high-temperature potential is the same as $V_0(\Phi)$ but with thermal masses added to $\mu_1^2$ and $\mu_2^2$. This is the lowest-order approximation used by Ref.~\cite{mjrmpatel}. 
%The critical temperature defined in this way is given by $V_0(\Phi_0) + V_1(\Phi_0, T_C) - \Delta V_A(\Phi_0, T_C) + \Delta V_B(\Phi(T_C),T_C)  = ...$. 
Since the potential is evaluated at two different minima, a gauge-invariant ring-improved $\Delta\phi$ is not well-defined and is not plotted in Fig.~\ref{fig:varxi}.

Ignoring thermal masses, one can see that the gauge-invariant critical temperature must be lower than the gauge-dependent critical temperature for any value of $\xi$: the gauge-dependent critical temperature is defined as the temperature at which $V(\Phi_{min}, T) = V(\Phi\!=\!0, T)$, but since $\Phi_{min}$ is the minimum of the potential, $V(\Phi_0,T) > V(\Phi_{min},T) = V(\Phi\!=\!0, T)$ and the gauge-invariant critical temperature must be lower. Conversely, the latent heat tends to be larger in the gauge-invariant method. The energy density decreases with increasing particle masses, so, as long as the masses are larger at $\Phi_0$ than at $\Phi_{min}$ (which is the case for weakly first-order transitions), the difference in energy densities between the symmetric and broken phases will be larger when evaluated at $\Phi_0$ than at $\Phi_{min}$. The addition of thermal masses tends to enhance both of these effects. 

The gauge-invariant method produces quite different results from the gauge-dependent calculation when the latter predicts a very weakly first-order transition. This is to be expected: the two methods perform calculations at very different field values when the gauge-dependent $\Delta\phi$ is small. When $g=1$ and the transition is more strongly first-order, the two methods agree much more closely. However, including thermal masses worsens the agreement. When $m_{gauge}/T \gtrsim 1$, higher-order terms in the effective potential dominate and using only the cubic term for ring-improvement is unreliable. The ring-improved gauge-invariant calculation should not be trusted in this case.

%We compare the gauge-dependent calculations to explicitly gauge-independent calculations (denoted by marks along the right side of each panel in Fig.~\ref{fig:varxi}) evaluated at the tree-level minima, where the ghost and goldstone degrees of freedom exactly cancel. This is the lowest-order approximation used by ref.~\cite{mjrmpatel}. The critical temperature defined in this way is given by $V(h_{tree},s_{tree},T_C) = V(h=s=0, T_C)$; $\Delta\phi$ is fixed at 246 GeV. When the gauge coupling is small and the transition ought to be weakly first-order, the gauge invariant critical temperature necessarily underestimates the transition temperature at one-loop (since $dV/dT$ is larger at larger field values), and similarly overestimates the difference in energy density. The gauge-independent and gauge-dependent calculations have as much as a factor of 10 disparity in their calculations of $\alpha$, with differences in $T_*$ accounting for another factor of $\sim$2 in the plotted $\alpha/T_*^4$. Neither method seems trustworthy in this scenario.

%At $g=1$, the two methods agree much more closely. The gauge-independent method yields a slightly higher transition temperature, both because it calculates the critical rather than nucleation temperature, and because, in this case, the one-loop zero-temperature corrections drive $\Delta\phi$ above its tree-level value even at finite temperature, so 

\begin{figure}[t]
\centering
\includegraphics[scale=1]{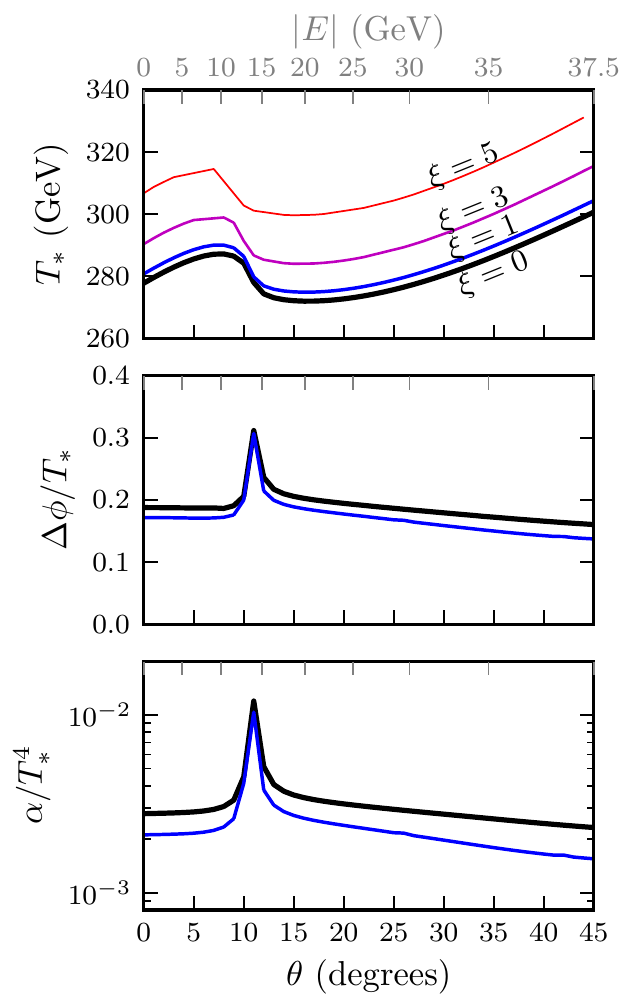}
\caption{\label{fig:thetadep1}
Gauge dependence in case 2 with $g=0.5$, but with $\theta$ (or equivalently $E$) varied and $\xi$ held fixed. The transition is second-order for $\xi \geq 3$ for all values of $\theta$.}
%Gauge dependence in case 2 with $g=0.5$, but with $\theta$ (or equivalently $E$) varied and $\xi$ held fixed at $\xi = 0$ (thick blue), 1 (purple) and 5 (thin red lines). The solid lines denote the primary transition which breaks the $h=0$ symmetry; the dashed lines denote a subsequent transition in the $h$-direction; and the dotted lines at low $\theta$ are for a transition along the $s$-direction. At $\xi=5$, the primary transition is second-order.}
\end{figure}

Fig.~\ref{fig:thetadep1} explicitly shows the (non-)importance of the cubic term ({\em via} $\theta$) in these scenarios. Regardless of whether the cubic term is large or small, the basic pattern of gauge-dependence is about the same. The phase transition grows more weakly first-order ($\alpha$ decreases) for increasing $\xi$ for all values of $\theta$. At $\xi \geq 3$, there is only a second-order transition. Note that the transition is most strongly first-order when $\tan\theta = \langle h_{tree} \rangle/\langle s_{tree} \rangle$, but still second-order for $\xi\geq 3$.

%Fig.~\ref{fig:thetadep1} explicitly shows the (non-)importance of the cubic term ({\em via} $\theta$) in this scenario. Regardless of whether the cubic term is large or small, the basic pattern of gauge-dependence is about the same. At low $\theta$, the first transition goes along the singlet direction (dotted lines in fig.~\ref{fig:thetadep1}, with two subsequent transitions primarily in the $h$-direction (solid and dashed lines). At higher $\theta$, the system transitions in the $h$-direction first, with a possible transition in the $s$-direction later. For all values of $\theta$, variations in $\xi$ greatly change the strength of the transition.

\subsection{Tree-level terms driving transitions}
\label{sec:cubicTrans}
Here we examine a scenario in which the cubic term is critical in determining the strength of the phase transition. Superficially, case 3 appears similar to case 2. Both have relatively large cubic terms, and both have the topology shown in Fig.~\ref{fig:treepot}e. However, a small change in model parameters can turn the saddle point in case 3 into a tree-level minimum (Fig.~\ref{fig:treepot}(\emph{d})), thus creating a potential barrier at zero temperature for which the tunneling rate may never be large enough to penetrate. Even without a tree-level barrier, the cubic term is sufficiently prominent to create a barrier at relatively low temperature: there is no barrier at $T=0$, but there is a barrier by $T \approx 100$ GeV for $\theta=60^\circ$. Slightly smaller values of $\theta$ decrease this temperature drastically.
The crucial distinction between cases 2 and 3 is that, even though both have large cubic terms, only in case 3 is the lowest eigevalue of the mass-squared matrix both negative and sufficiently small in magnitude
%does the mass-squared matrix  lowest (negative) eigenvalue with sufficiently small magnitude
%small enough negative value of its mass-squared matrix at the origin  
(i.e., a small negative value of $\mu_1^2$)  that the origin can become a minimum at low temperature with the cubic term providing the separation between the symmetric and broken phases.

 \begin{figure}[t]
\centering
\includegraphics[scale=1]{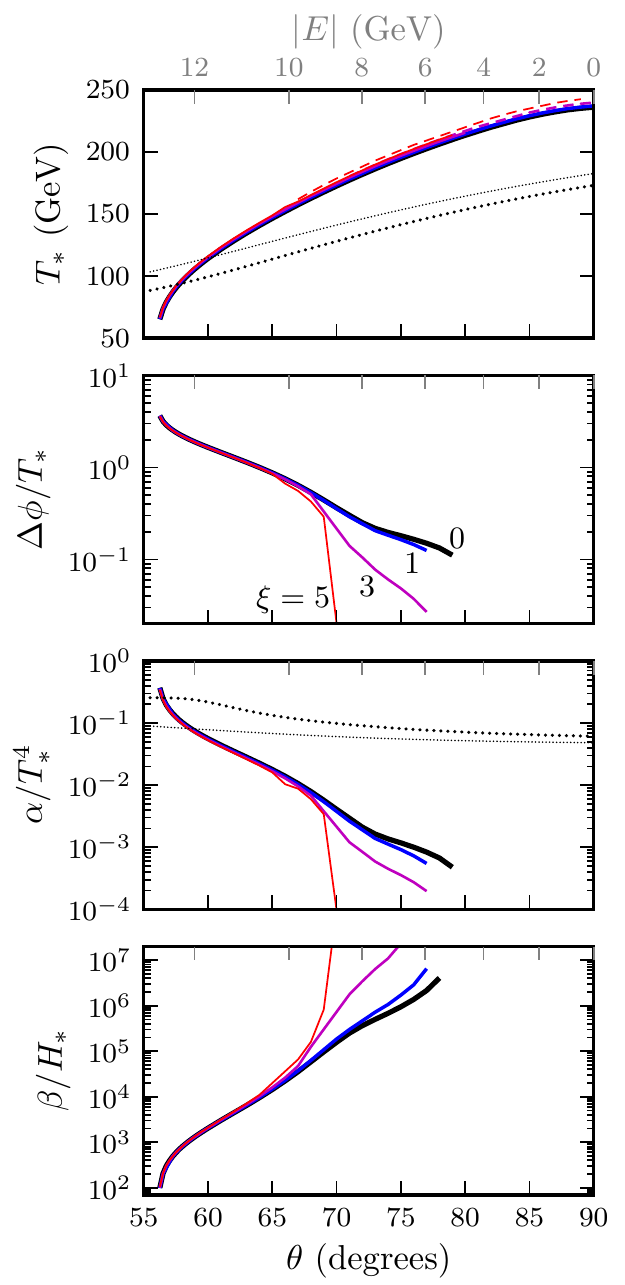}
\caption{\label{fig:thetadep2}
Similar to Fig.~\ref{fig:thetadep1}, but for case 3 with $g=0.3$. 
At $\theta \gtrsim 70^\circ$, the symmetry-breaking transition is second-order for $\xi = 3$ and 5 (dashed lines), followed by a weakly first-order transition. At $\theta \gtrsim 80^\circ$, the symmetry-breaking transition is second order for all plotted values of $\xi$ and the first-order transitions have disappeared. The thick/thin black dotted lines show the gauge-invariant critical temperature and latent heat calculations with/without thermal mass corrections.}
%At $\theta \gtrsim 70^\circ$, the symmetry-breaking transition precedes a separate weakly first-order transition. For high enough $\theta$, the initial symmetry-breaking is second-order for all plotted values of $\xi$. }
\end{figure}

Since the phase transition is strongly first-order even for $g=0$ (that is, without any gauge bosons at all), the gauge dependence is not nearly as severe as in cases 1 and 2. 
It is still present though. For example, at $\theta=60^\circ$, $T_*$ increases by 1.6\% from $\xi=0$ to $\xi=5$, and $\alpha$ increases by only 0.1\%.
%It is still present though. For example, $\alpha/T_*^4$ increases by about 30\% from $\xi=0$ to $\xi=5$ when $g=0.3$ and $\theta = 60^\circ$.

By increasing $\theta$ from $55^\circ$ up to $90^\circ$, one moves successively through topologies (\emph{d}), (\emph{e}), (\emph{f}), and (\emph{a}) in Fig.~\ref{fig:treepot}. At around $\theta=70^\circ$ and $|E|=9$ GeV, the cubic term is small enough so that a first-order phase transition requires $g>0$. At this point, the gauge dependence becomes much more obvious, as is seen in Fig.~\ref{fig:thetadep2}. At high enough $\theta$ and low enough $E$, the symmetry-breaking transition is second-order for all plotted values of $\xi$.%, followed by a subsequent weakly first-order transition. Note that for $\xi=0$ and $\xi=1$, the first transition at high-$\theta$ is continuous with the transition at low-$\theta$, whereas for $\xi=5$ the secondary transition is continuous with the transition at low-$\theta$.

\section{Discussion and conclusions}
\label{sec:concl}

The main conclusion of the present study is that the inclusion of a singlet scalar degree of freedom does not generally alleviate the gauge-dependence problem in the electroweak phase transition, even when it has large couplings to the Higgs. Moreover, a significant, tree-level cubic singlet-Higgs interaction  does not in itself guarantee a strongly first-order phase transition. On the other hand, when the phase transition is strongly first order, the gauge-dependence appears to be less pronounced than in the generic case. 
Such a situation occurs either when the gauge coupling is relatively large or when the tree-level singlet-Higgs cubic term acts in concert with small negative mass-squared values at $h=s=0$ to create a potential barrier at low temperature. Otherwise, when the phase transition is only weakly first-order, or borderline strongly first-order, the gauge-dependence can be drastic regardless of the presence of a cubic term. This dependence may change not only the strength of the phase transition, but also its overall character. In such circumstances, one cannot make a gauge-independent determination of whether the transition is first or second-order, nor can one even determine whether or not the transition comes from a symmetry-preserving vacuum. The explicitly gauge-independent calculation of the critical temperature $T_C$ using the lowest-order result in Ref.~\cite{mjrmpatel} can give a rough estimate of the transition temperature $T_*$, but only when the amount of super-cooling is small\footnote{In this case, the onset of nucleation occurs for $T$ very close to $T_C$}, which is hard to achieve with a tree-level barrier. A similar gauge-independent calculation of $\alpha$ is only reasonable when the transition is already known to be strongly first-order.

As emphasized in our earlier work \cite{Wainwright:2011qy}, the appearance of gauge-dependence in physical quantities such as $T_*$, $\alpha$, and $\beta$ should engender caution when attempting to draw phenomenological conclusions from computations performed in a specific gauge. In the ideal situation, a gauge-invariant computation using non-perturbative methods would be used to explore various Standard Model extensions that may lead to a first order electroweak phase transition, though a comprehensive exploration is at present prohibitively expensive. In the meantime, various gauge-invariant perturbative techniques, such as the loop expansion~\cite{mjrmpatel} or Hamiltonian formulation~\cite{boyan}, may at least point to regions of parameter space in a given model where transitions of different character occur. If, as we find for the Abelian Higgs plus singlet model (and as, perhaps, maybe speculatively be a more general pattern) the gauge-dependence of conventional perturbative computations is mitigated by a strongly first order transition triggered by gauge-independent terms, one might expect to find rough agreement with the results of manifestly gauge-invariant analyses. The present study shows, however, that such a conclusion should be carefully qualified on a case-by-case basis.

\section*{Acknowledgements}
SP is partly supported by an Outstanding Junior Investigator Award from the US Department of Energy and by Contract DE-FG02-04ER41268, and by NSF Grant PHY-0757911. CW acknowledges support from the NSF graduate research fellowship program. MJRM was supported in part by US Department of Energy Contract DE-FG02-08ER41531 and the Wisconsin Alumni Research Foundation.

\end{document}